\documentclass[slac_one]{revtex4}
\usepackage{graphicx}
\usepackage{fancyhdr}
\begin{document}

\title{{\small{Hadron Collider Physics Symposium (HCP2008),
Galena, Illinois, USA}}\\ 
\vspace{12pt}
Top Quark Properties} 

%

\author{Catalin I. Ciobanu \\ For the CDF and D\O ~Collaborations}
\affiliation{{\it IN2P3-CNRS LPNHE Laboratory and Pierre \& Marie  Curie University\\
4 Place Jussieu, 75252 Paris Cedex 05 France}}

\begin{abstract}
In this document we present several recent (July 2008) results from studies of the 
top quark properties at the CDF and D\O~ experiments at the Tevatron. Measurements
of several top quark properties, as well as tests of new physics in the top quark production
and decay sectors are presented. In the latter case, no significant evidence for physics
beyond the Standard Model is uncovered, and the tightest constraints to date are placed on
most of the new physics scenarios investigated.
\end{abstract}

\maketitle

\thispagestyle{fancy}


\section{Introduction} 
According to the Standard Model (SM), in $p\bar{p}$ collisions at the Tevatron top
quarks can be created in pairs via the strong force, or singly via the
electroweak interaction. Due to its higher cross-section and  
smaller associated backgrounds, the strong production ($t\bar{t}$) represents the main
channel used for performing detailed measurements of the properties of the top quark. 
In terms of its decay, a top quark is expected to disintegrate into a $W$ boson and
a $b$-quark almost 100\% of the time. Thus, the top candidate events can be classified 
based on the subsequent decays of the $W$ daughter into: a) ``dilepton'' events if both $W$
bosons decay leptonically, b)  ``lepton+jets'' events if one $W$ decays leptonically and the
other one hadronically, and c) ``all hadronic'' events if both $W$'s decay hadronically. 
The last category of top candidates (all hadronic) are typically dominated by 
multijet background contributions, and will not be used in the results described here. 
In the following sections, we will present the results from top charge measurement, 
as well as measurements related to the production and the decay of top quarks.

\section{Top Properties: Charge}
While the top charge is expected to be $+\frac{2}{3}e$, there are models
which include an exotic $-\frac{4}{3}e$ top quark that couples to a 
right-handed $b$ quark \cite{chgtheo} and decays into a negative $W$ boson and
a $b$ quark. 
Both the CDF and D\O~ experiments have performed measurements to test if the top 
charge is  $\frac{2}{3}e$ or $-\frac{4}{3}e$. 
The challenge is therefore to reconstruct the charges of the $W$ bosons and the $b$ quarks
in the event, and subsequently pair the $W$ and the $b$'s to reconstruct the top
decay chain. In the CDF analysis, dilepton and lepton+jets top candidate events are
selected by first requiring evidence for the leptonic decay of a $W$: an electron (muon) 
with large transverse energy $E_{T}$($p_{T}$)$>$20 GeV and large transverse missing
energy from the undetected neutrino:  ${E_T\!\!\!\!\!\!\!/ \quad}>$20 GeV. For the
lepton+jets subsample, three jets with $E_{T}>$20 GeV and a fourth jet with
 $E_{T}>$12 GeV are required. To further reduce background contamination, at least
two of these jets are required to be identified (tagged) as $b$-jets using displaced 
vertex information as measured by the CDF silicon vertex detector. In the case of the
dilepton subsample, the presence of another high transverse momentum, opposite-sign 
lepton is demanded, along with the presence of at least two jets with $E_{T}>$15 GeV.
Of the two jets, one or both should have been tagged as $b$-jets by the silicon 
detector. A minimum threshold of 200 GeV on the total transverse energy in the event 
(scalar-sum including  $E_T\!\!\!\!\!\!\!/ \quad$) is also imposed.

The charge of the $b$-jets is determined by performing a $p_{J}$-weighted
\footnote{ $p_{J}$ is the projection of track's momentum along the jet axis.} 
sum of the charges of the tracks found within a cone of radius  
$\Delta R=\sqrt{\Delta \Phi ^{2} + \Delta \eta ^{2}}=$0.4 around the jet axis.
A negative (positive) weighted sum indicates a $b$ ($\bar{b}$) quark. For real $b$-jets 
present in CDF data this method have been shown to determine the correct charge 
roughly 60\% of the time. 

Finally, the pairing of $W$ and $b$ in dilepton events is performed by computing the 
invariant mass of the lepton and $b$-jet ($M_{\ell b}$). This mass can be abnormally high for 
the wrong combinations; the CDF pairing exploits this fact, first by requiring one 
combination over a certain threshold and then choosing the opposite $\ell-b$ pairing. 
This procedure was estimated to have a 95\% success rate, with an efficiency of 39\%.
In lepton+jets events, a kinematic fitter $\chi^{2}$ using reconstructed top and
$W$ mass constraints is calculated and the lowest $\chi^{2}$ combination is retained.
The correct pairing is found 86\% of the time, with an efficiency of 53\%.

\begin{figure*}[tbh]
\centering
\includegraphics[width=100mm]{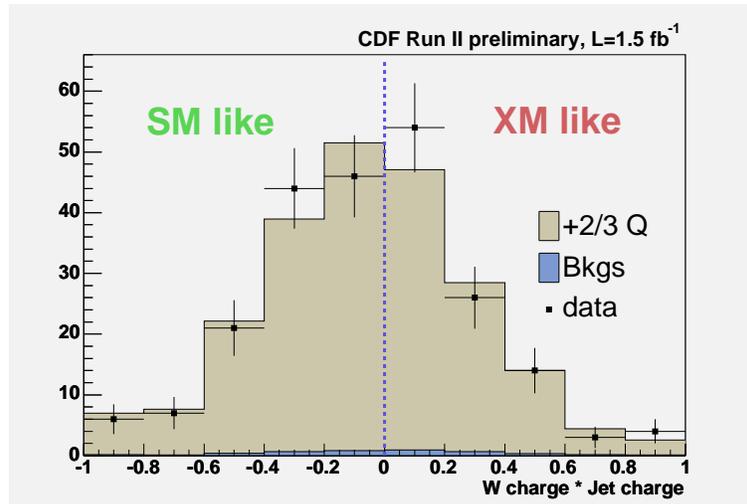}
\caption{CDF: The distribution of the product between the $W$ (lepton) charge and the jet charge. The 
histograms show the Standard Model expectation, while the points represent the CDF data.} \label{chgplot}
\end{figure*}

The results of this analysis are presented in Figure \ref{chgplot}. These results favour
the SM hypothesis over the $-\frac{4}{3}e$ top charge hypothesis. The latter hypothesis is
rejected at 87\% confidence level (C.L.). We note that a top charge measurement has previously
been performed by the D\O~ Collaboration \cite{dcharge} with similar conclusions.

\section{Top Quark Production}
\subsection{Top Quark Production Mechanism: $q\bar{q}$ versus $gg$ initial state}
At the Fermilab Tevatron, the top pairs are expected to originate
predominantly from a 
$q\bar{q}$ initial state: $\sigma(gg\rightarrow t\bar{t} X) / 
\sigma(q\bar{q}\rightarrow t\bar{t} X)\approx 15$\% $/$ 85\%, with somewhat
large (${\cal{O}}(10$\%))uncertainties.

\begin{figure*}[tbh]
\centering
\includegraphics[height=60mm,width=80mm]{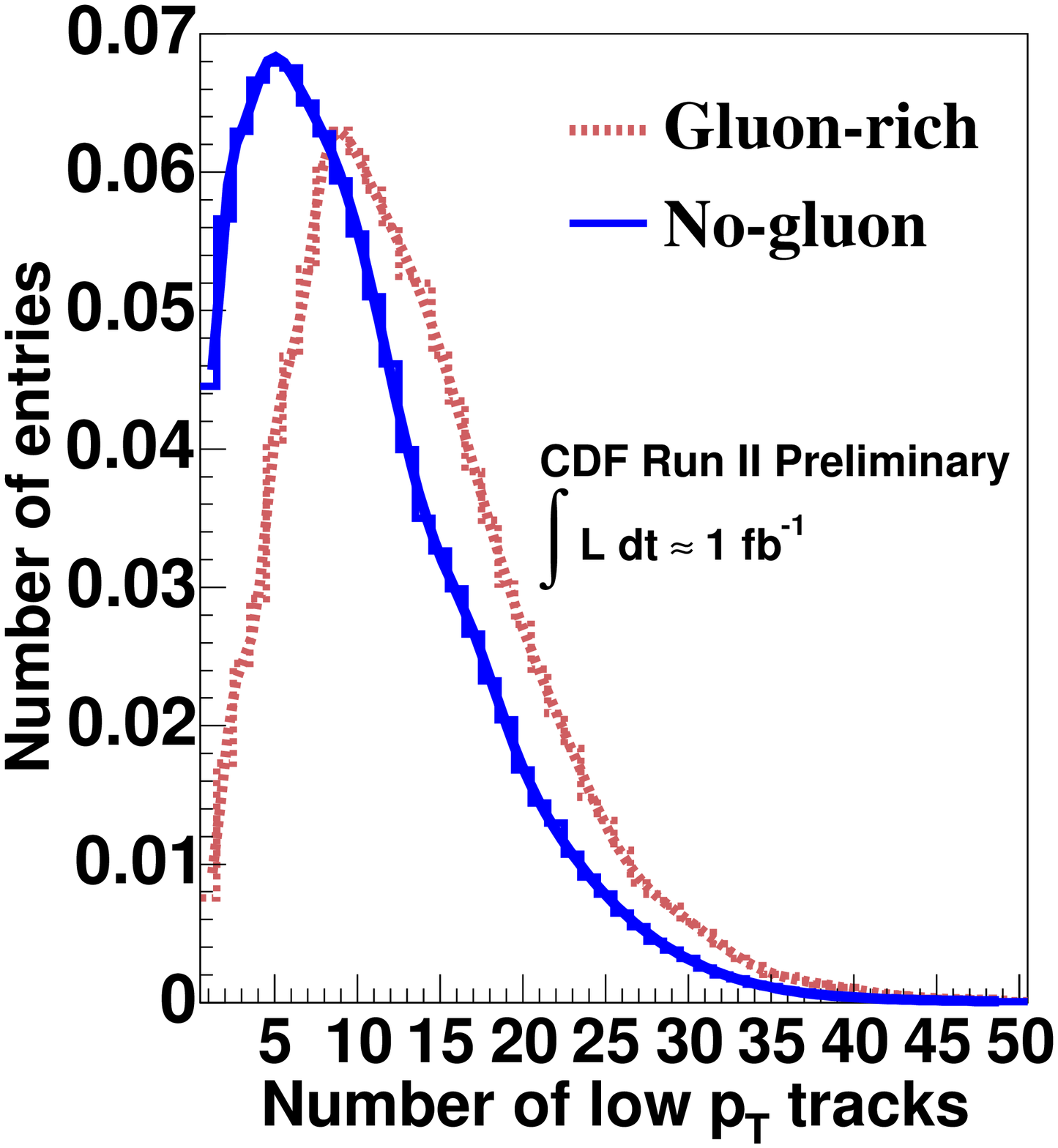}
\includegraphics[height=60mm,width=80mm]{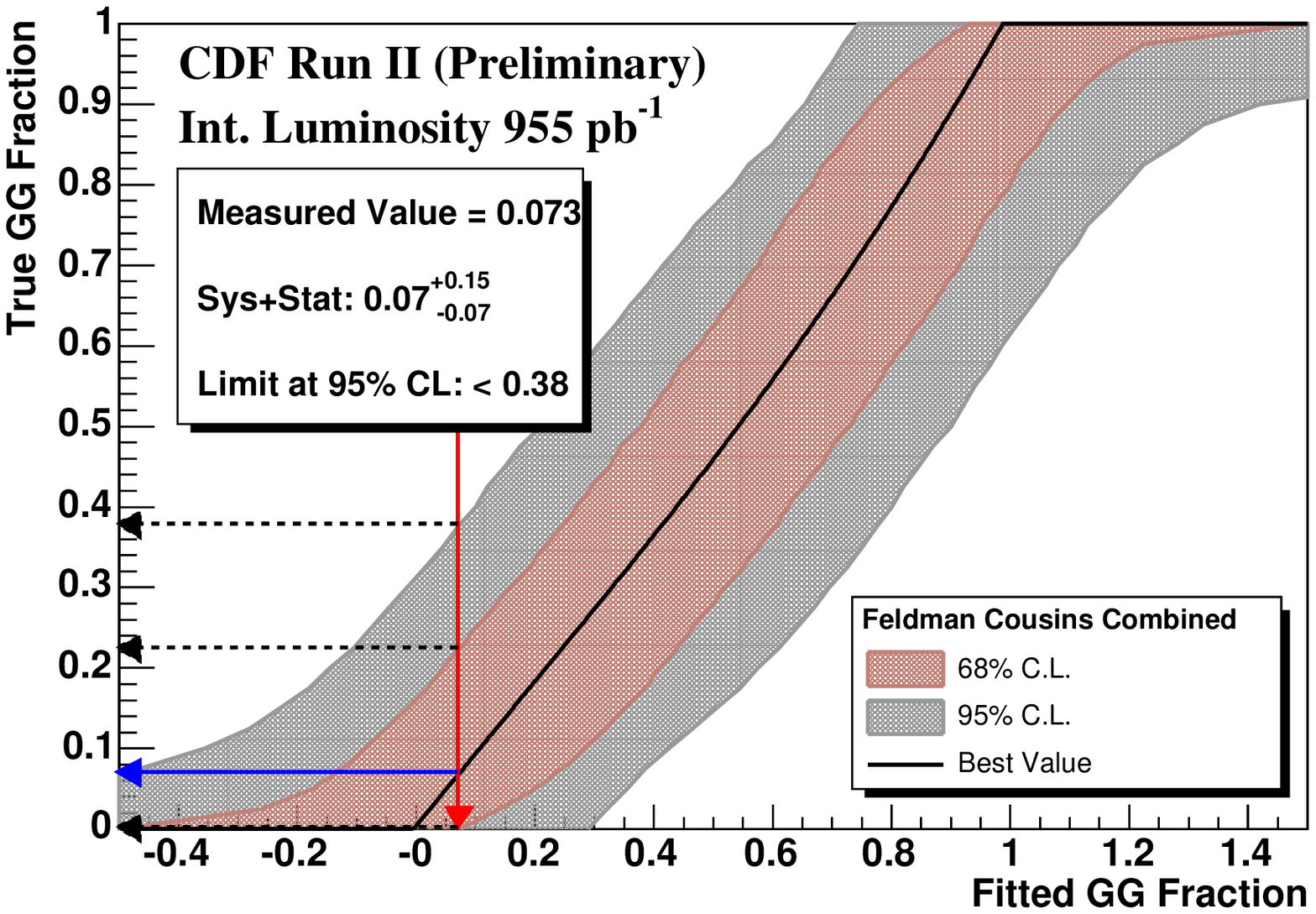}\caption{Left plot: CDF: 
The distributions of the low-$p_{T}$ track multiplicity in the no-gluon and the gluon-rich cases.
Right plot: CDF top production mechanism combination: Feldman-Cousins 
$\sigma(gg\rightarrow t\bar{t} X) / [\sigma(q\bar{q}\rightarrow t\bar{t} X)+\sigma(gg\rightarrow t\bar{t} X)]$
plot showing the central value and the 68\% and 95\% C.L. intervals.
} \label{combo}
\end{figure*}

Two CDF measurements \cite{gg} attempt to measure the production cross section ratio. The first 
measurement takes advantage of the large correlation between the average number of gluons and 
the average number of low $p_{T}$ charged particles present in a given sample. 
The $W$+0jet sample and the dijet (80-100GeV) sample are then used to extract the distributions
of the low-$p_{T}$ track multiplicity in the ``no-gluon" case, and the "gluon-rich" case,
respectively (Figure \ref{combo}). 
These two templates are used in a simple maximum likelihood fit to derive the fractions of
the no-gluon and the gluon-rich events present in the CDF data. After background subtraction
is performed, it is found that
$\sigma(gg\rightarrow t\bar{t} X) / \sigma(q\bar{q}\rightarrow t\bar{t} X) = 
0.07\pm 0.14$(stat)$\pm 0.07$(syst). 

The other CDF measurement of the production mechanism relies on an Artificial Neural Network (ANN) designed to
separate events originating from $q\bar{q}$ and $gg$ initial state. The ANN uses eight variables, of which two 
(six) variables characterize the production (decay) of the top pair. A maximum likelihood fit yields
$\sigma(gg\rightarrow t\bar{t} X) / [\sigma(q\bar{q}\rightarrow t\bar{t} X)+\sigma(gg\rightarrow t\bar{t} X)] =
-0.075$, or 
$\sigma(gg\rightarrow t\bar{t} X) / [\sigma(q\bar{q}\rightarrow t\bar{t} X)+\sigma(gg\rightarrow t\bar{t} X)] <0.61$
at 95\% C.L. The two measurements are combined to obtain
$\sigma(gg\rightarrow t\bar{t} X) / [\sigma(q\bar{q}\rightarrow t\bar{t} X)+\sigma(gg\rightarrow t\bar{t} X)]
= 0.07^{+0.15}_{-0.07}$ (stat+syst) (see Figure \ref{combo}). 

\subsection{Forward-Backward Asymmetry in Top Production}
The D\O~Collaboration has recently published the first measurement of the 
forward-backward asymmetry in top production \cite{d_afb}. While a small
asymmetry of the order of 5\% is expected to arise from QCD calculations
\cite{afb_SM}, several new physics phenomena could drastically increase
this asymmetry \cite{afb_BSM}.  The asymmetry $A_{FB}$ is defined as 
the count ratio $(N_{F}-N_{B})/(N_{F}+N_{B})$, where an event is counted as 
forward (backward) if the rapidity difference $\Delta y = y_{t}-y_{\bar{t}}$ 
is positive (negative). This measurement is performed in the reconstructed
$t\bar{t}$ rest frame. The results are shown in Table \ref{dtab}. 
While the values of the asymmetry are
somewhat higher than their expected values (a few percent, depending on jet channel) 
they also have sizable uncertainties, and are therefore in statistical agreement 
with these expectations.

\begin{table}[tbh]
\begin{center}
\caption{D\O: Number of selected events and fit results in the data.}
\begin{tabular}{lccc}
\hline \hline
 \hspace{1cm} & $\geq4$ jets  \hspace{1cm} & 4 jets  \hspace{1cm}  & $\geq5$ jets  \\
\hline
No. Events  \hspace{1cm} & 376  \hspace{1cm} &  308  \hspace{1cm}  & 68 \\
\hline 
$t\bar{t}+X$ \hspace{1cm}& 266$^{+23}_{-22}$  \hspace{1cm} & 214$\pm$20  \hspace{1cm} & 54$^{+10}_{-12}$ \\
$W$+jets     \hspace{1cm} & 70$\pm$21  \hspace{1cm} & 61$^{+19}_{-18}$   \hspace{1cm} & 7$^{+11}_{-5}$ \\
Multijets    \hspace{1cm} & 40$\pm$4   \hspace{1cm} & 32.7$^{+3.5}_{-3.3}$  \hspace{1cm} & 7.1$^{+1.6}_{-1.5}$ \\
$A_{FB}$     \hspace{1cm} & (12$\pm$8)\%  \hspace{1cm} & (19$\pm$9)\%  \hspace{1cm} & (-16$^{+15}_{-17}$)\%  \\
\hline \hline
\end{tabular}
\label{dtab}
\end{center}
\end{table}

A similar search has recently been performed by the CDF Collaboration \cite{jeanny}. The $A_{FB}$ was
measured in both the reconstructed rest frame of the $t\bar{t}$ system and in the 
laboratory frame, and corrected to the intrinsic parton-level value to allow an
easy comparison between the experimental $A_{FB}$ result and the 
theoretical predictions. 
The results \footnote{The expected relationship between the laboratory frame and 
the $t\bar{t}$ frame asymmetries is: $A_{FB}^{t\bar{t}}\sim 1.3\cdot A_{FB}^{p\bar{p}}$.}
$A_{FB}^{t\bar{t}}=0.24\pm0.13$(stat)$\pm0.04$(syst) and 
$A_{FB}^{p\bar{p}}=0.17\pm0.07$(stat)$\pm0.04$(syst)
are higher than, but consistent with the theoretical SM calculation.

\subsection{Resonance Searches}

\subsubsection{$t\bar{t}$ Resonance Searches}
The D\O~and CDF Collaborations have performed searches for narrow-width, high-mass particles X decaying to 
top quark pairs. Such particles are predicted in many extensions of the SM, such as extended gauge 
theories \cite{E6}, Kaluza-Klein excited states of gluons or $Z$ bosons \cite{KK}, and topcolor \cite{topcolor}
to name a few. 

The D\O~search uses the lepton+jets subsample. At least three high-$E_{T}$ jets are required to be
present in every event; at least one of these jets should be identified as a $b$-jet using 
a neural networks tagging algorithm \cite{d_nnb}. The four momenta of the jets, lepton, and 
neutrino \footnote{The longitudinal, or $z$-component of the neutrino momentum is obtained 
from the $W$ mass constraint $M_{\ell \nu}=M_W$. This condition yields a quadratic equation 
in $p_{z}^{\nu}$; if there are two real solutions, the smaller $|p_{z}|$ one is chosen. If there
are no real solutions, then $p_{z}$ is set to zero.} are used
to reconstruct the invariant mass of the $t\bar{t}$ system.  
  
\begin{figure*}[tbh]
\centering
\includegraphics[width=80mm]{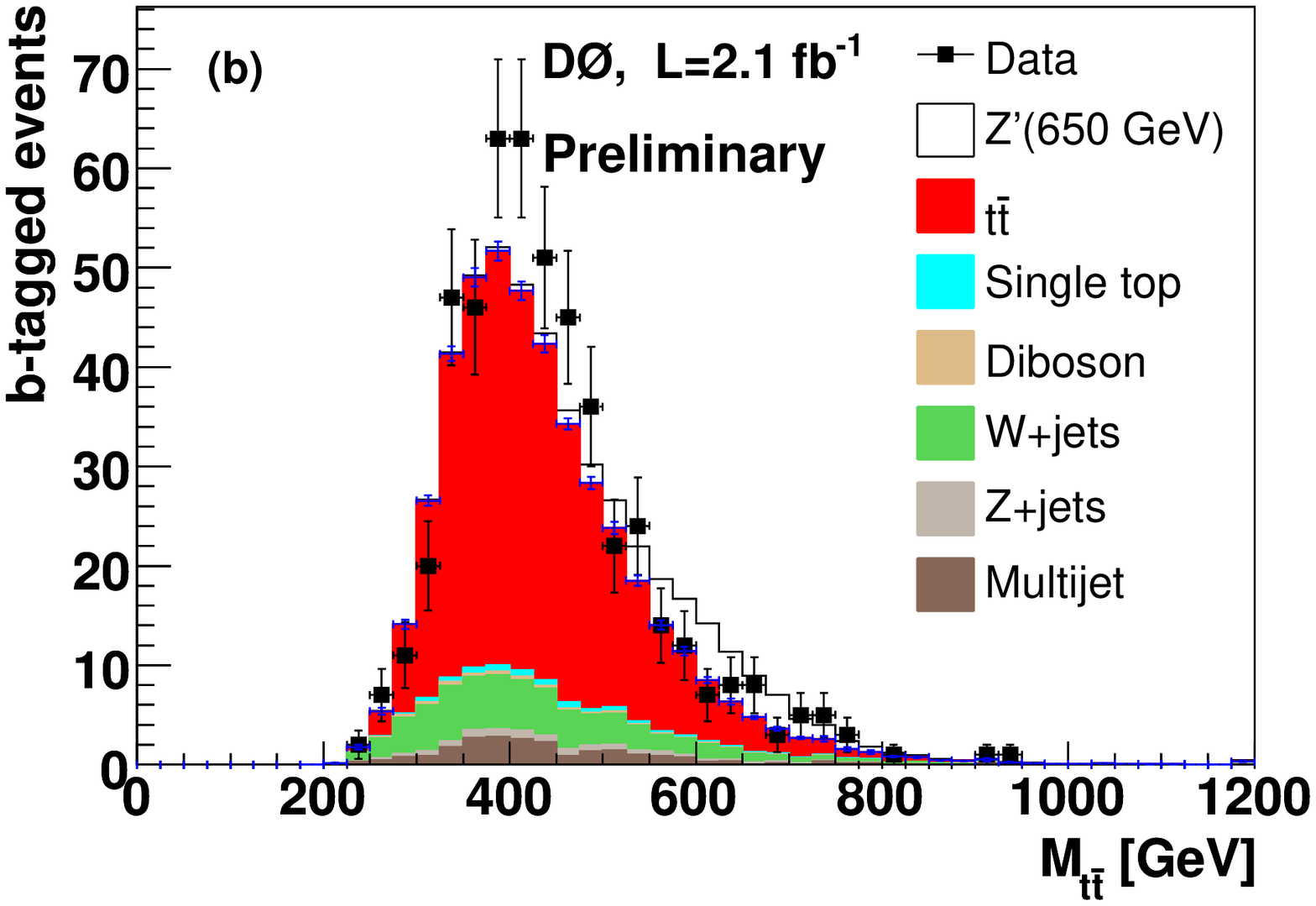}
\includegraphics[width=80mm]{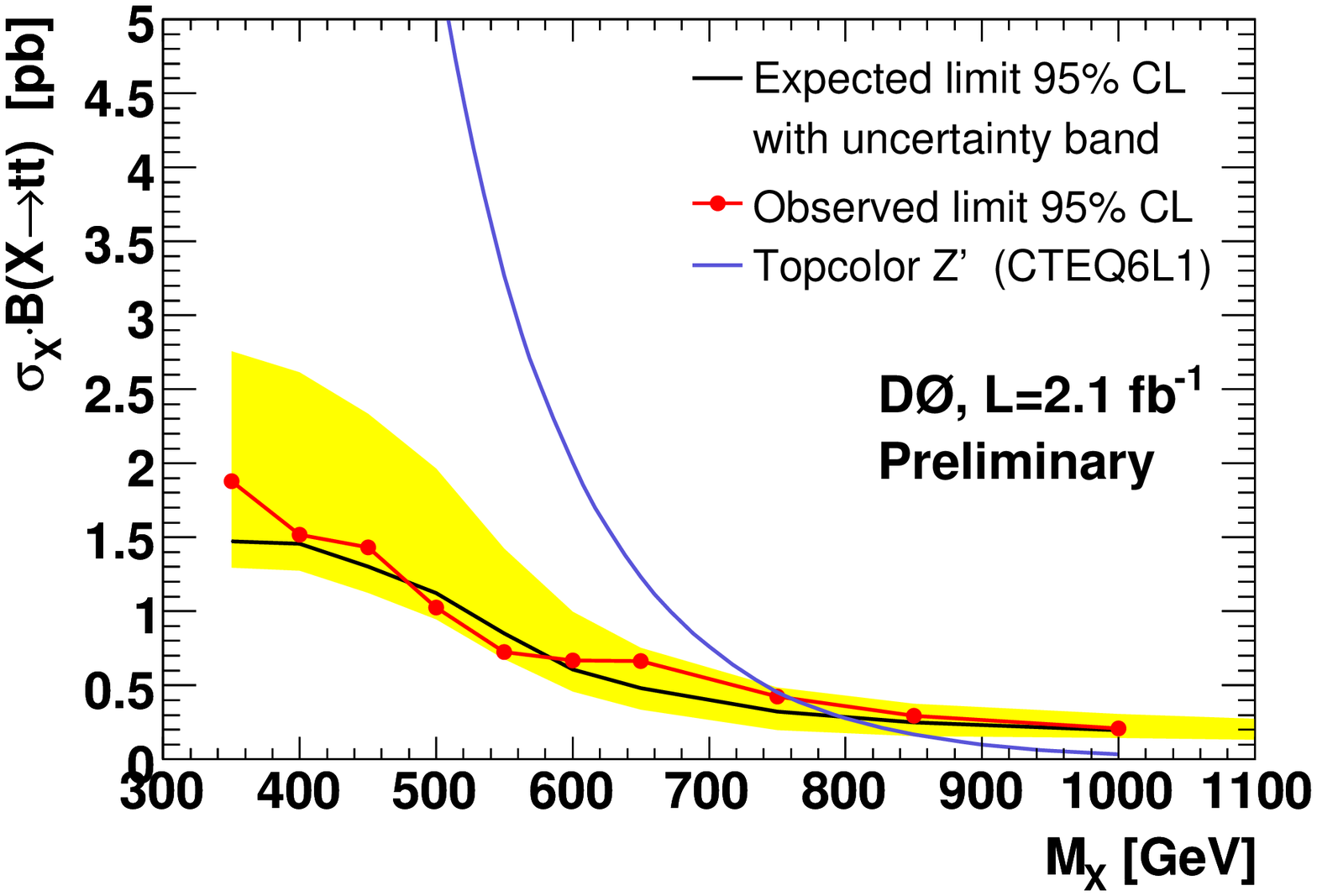}
\caption{Left plot: D\O: The reconstructed $M_{t\bar{t}}$ spectrum in the $\geq$4 jets channel, showing the data (points) and the
SM contributions (color histograms). A 650 GeV top-color-assisted model is shown by the white (open)
contributions. Right plot: D\O: 95\% C.L. exclusion limits on $\sigma_X \cdot BR(X\rightarrow t\bar{t})$. } \label{d_restt}
\end{figure*}

The left plot of Figure \ref{d_restt} shows the distribution of the reconstructed top pair 
mass $M_{t\bar{t}}$ in the lepton+$\geq$4 jets channel for data (points) and SM expectation
(colored histograms). A 650 GeV top-color-assisted signal \cite{topcolor} is shown
by the white contribution stacked on top of the SM expectation (open histogram). The data 
are found to agree well with the SM expectation. 95 \% C.L. limits are set on 
$\sigma_X \cdot BR(X\rightarrow t\bar{t})$ for different $M_X$ values, as shown in 
the right plot of Figure \ref{d_restt}. 
One can see that a leptophobic $Z^{\prime}$ with a width 
$\Gamma_{Z^{\prime}}=1.2$\%$M_{Z^{\prime}}$ is excluded (95\% C.L.) up to a mass of 760 GeV.

\begin{figure*}[t]
\centering
\includegraphics[width=100mm]{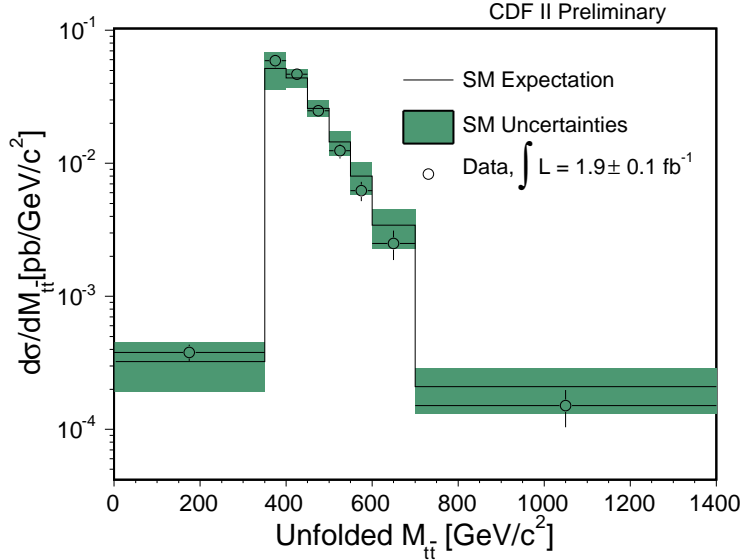}
\caption{ CDF: The differential top pair cross section in bins of $M_{t\bar{t}}$.
The circles represent CDF data, while the histogram represents the expected
SM contribution with uncertainties (luminosity uncertainty is not included).} 
\label{alis}
\end{figure*}

A similar search at CDF observes no significant evidence for new particles decaying to $t\bar{t}$. Using the 
lepton+$\geq 4$ jets sample, the CDF Collaboration
has also measured the differential top pair production cross-section in bins of $M_{t\bar{t}}$. The result is shown
in Figure \ref{alis} where a good agreement between CDF data (points) and expected distribution (histogram) is
apparent. To quantify this agreement, a ``p-value", defined as the fraction of SM simulated experiments which 
look as discrepant as the data or worse, is calculated; the result of 0.45 confirms no excess over SM prediction
is observed.    

\subsubsection{Other Resonance Searches}

Other resonance searches in the top sector at the Tevatron include $t^{\prime}$ searches
and scalar top searches. The CDF Collaboration has performed a search for a hypothetical new quark, 
$t^{\prime}$, whose decay features the same final state particles as the SM top. 
The $t^{\prime}$ particle could belong to a fourth quark generation in which the mass of the $b^{\prime}$ 
partner is large enough to satify: $M_{b^{\prime}} > M_{t^{\prime}} - M_{W}$; in this case, 
$t^{\prime}$ would decay predominantly into $Wq$ ($q=d,s,b$). 
The variables used to discriminate the $t^{\prime}$ signal from the associated 
background (which includes SM $t\bar{t}$ production) are the $t^{\prime}$ 
reconstructed mass $M_{reco}$ and the total transverse energy in the event $H_{T}$.
A simultaneous analysis of these variables excludes the presence of a 
$t^{\prime}$ signal up to a mass of 284 GeV (at 95\% C.L.).

Searches for scalar top quarks (stop) have been performed by both the CDF and D\O~Collaborations.
Because of the large top mass, the mixing between the stop eigenstates can be substantial, and the
lighter stop $\tilde{t_{1}}$ could be lighter than the top quark \cite{marcela}. 
The D\O~search focuses on this scenario in which $\tilde{t_{1}} \bar{\tilde{t_{1}}}$ 
pairs are produced and subsequently decay as 
$\tilde{t_{1}}\rightarrow \tilde{\chi_{1}}^{+} b \rightarrow W^{+} \tilde{\chi_{1}}^{0} b$.  
The final state therefore resembles that of $t\bar{t}$ production, with some additional
missing energy from the neutralinos. Multivariate likelihood discriminants are built 
for six models, defined by $M_{\tilde{t_{1}}}$ and $M_{\tilde{\chi_{1}}^{\pm}}$.
In all cases, no significant evidence of a signal is found, and upper limits are set on
the cross section of these models, ranging from 7-12 times the theoretical calculations
(at 95\% C.L.). The CDF analysis \cite{andrew} draws similar conclusions.

\section{Top Quark Decay}
\subsection{$W$ helicity in Top Decay}
\begin{figure*}[tbh]
\centering
\includegraphics[width=80mm]{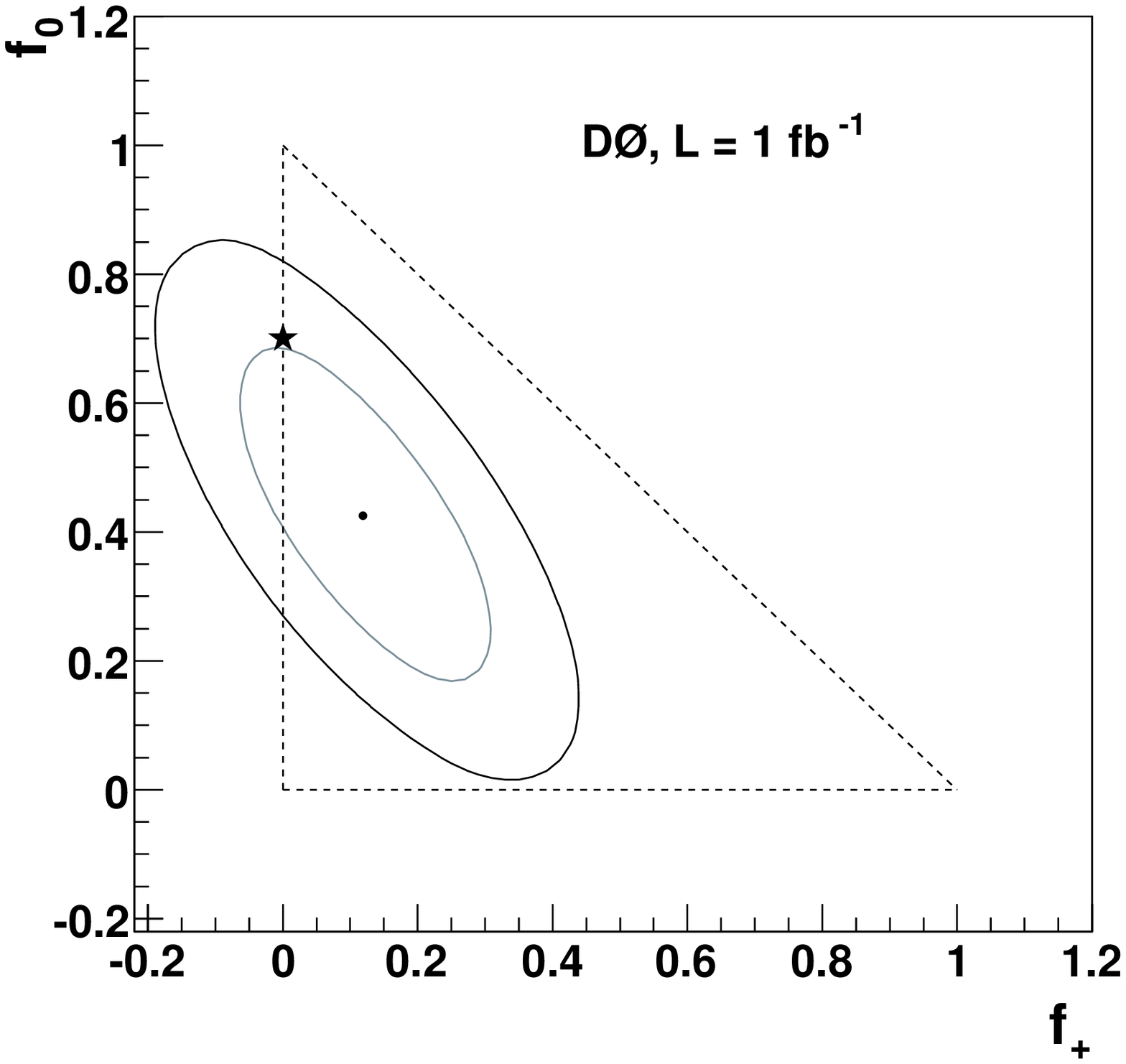}
\includegraphics[height=80mm]{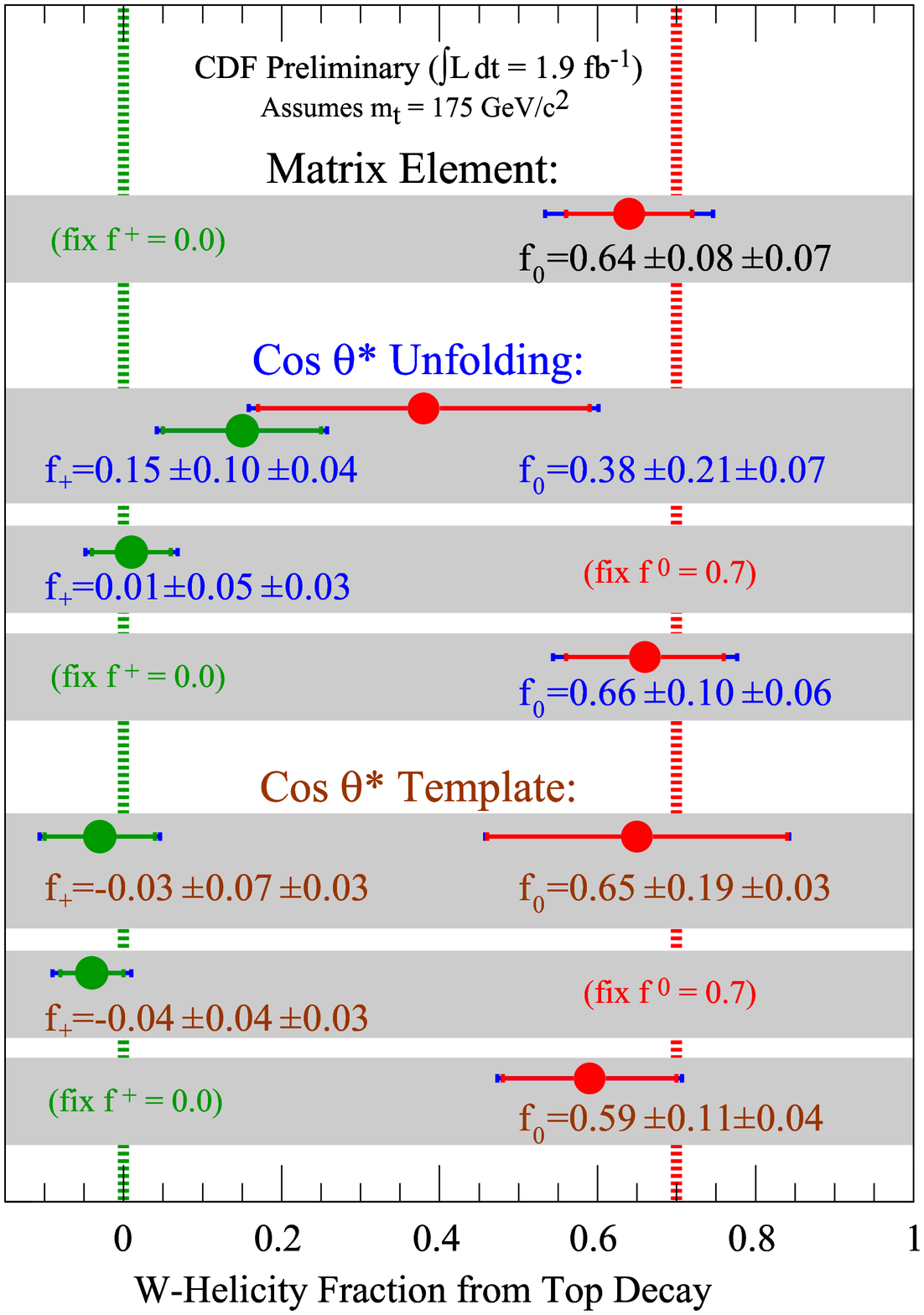}
\caption{Left plot: D\O: The simultaneous fit for the fractions of longitudinal and
right-handed $W$ bosons from top decays. The
ellipses are the 68\% and 95\% C.L. contours, the triangle borders the
physically allowed region where $F_{0}$ and $F_{+}$ sum to one or less, and
the star denotes the SM values. Right plot: CDF results from the three 
individual $W$ helicity analyses.} 
\label{heli}
\end{figure*}

The V$-$A structure of the weak current $t\rightarrow Wb$ requires the following 
helicity fractions $F_{0}=0.7$, $F_{-}=0.3$, and $F_{+}=0.0$ for the longitudinal,
left-handed, and right-handed $W$ contributions, respectively. An  $F_{+}$ value 
significantly different from zero would indicate the presence of physics beyond the SM.
The D\O~Collaboration has recently published \cite{d_heli} results from the measurement 
of the fractions of $W$ helicity contributions using the $\cos \theta$* variable. 
\footnote{ $\theta$* is the angle between the top quark and the lepton 
from $W$ decay, measured in the $W$ boson rest frame.} 
Both the dilepton and the lepton+jets subsamples are used for this measurement. The
$F_{0}$ and $F_{+}$ fractions are fitted simultaneously, while  $F_{-}$ is constrained 
to $1-F_{0}-F_{+}$. The result of the fit is shown in Figure \ref{heli}:
$F_{0} = 0.425 \pm 0.166$(stat)$\pm 0.102$(syst) and 
$F_{+} = 0.119 \pm 0.090$(stat)$\pm 0.053$(syst)

The CDF Collaboration has employed a three-prong approach for performing this 
measurement. The first two methods rely on the cosine of the decay angle $\theta$* of 
the charged lepton in the $W$ rest frame measured with respect to the direction of 
motion of the $W$ boson in the top-quark rest-frame. Fitting for both 
$F_{0}$ and $F_{+}$ fractions simultaneously and combining the two analyses, 
CDF obtains $F_{0}=0.66 \pm 0.16$(stat+syst) and $F_{+}=-0.03\pm 0.07$(stat+syst). 
The third method relies on the matrix element technique previously developed in 
Run I \cite{flo}. In this measurement the $F_{+}$ and $F_{-}$ fractions are 
fixed to zero and 1-$F_{0}$, respectively. The method yields 
$F_{0}=0.637\pm 0.084$(stat)$\pm 0.069$(syst). The results from the three CDF 
methods are summarized in Figure \ref{heli}.

To summarize, the D\O~and CDF results are compatible with each other and with the SM
expectations. More data will be needed for a conclusive determination of the three
polarization fractions.

\subsection{Suppressed Top Decays: $t\rightarrow Zq$}
In the SM, the flavor changing neutral current (FCNC) $t\rightarrow Zq$ is
predicted to occur at a tiny rate ${\cal O}(10^{-14})$; however, several scenarios involving 
physics beyond the SM could enhance this branching ratio up to values of the order ${\cal O}(10^{-4})$. 
Thus, any sign of this rare decay at the Tevatron would be an indication of new physics. 

A CDF search for $t\rightarrow Zq$ decays is based on a template fit to a $\chi ^{2}$ variable using kinematic
constraints present in the top FCNC events; one of these constraints is that the reconstructed mass of the
$Zq$ system approach the top mass, within a small window. The subsample used for this study is the 
$Z+\geq$4 jets, where the $Z$ decayed to a pair of oppositely-charged electrons or muons. 
The analysis of the kinematic $\chi ^{2}$ reveals no significant evidence for a signal beyond the SM. 
A Feldman-Cousins approach yields the tightest constraint to date in this top decay channel:
$BR(t\rightarrow Zq)<3.7$\% at 95\% C.L.

\subsection{Exotic Top Decays: $t\rightarrow H^{+}b$}
Charged Higgs bosons are predicted in many extensions of the SM, such as supersymmetry or grand unified symmetries
If these bosons have massed below that of the top quark, the process $t\rightarrow H^{+}b$ would lead to a 
decrease in the SM branching ratio BR($t\rightarrow W^{+}b$) assumed to be very close to unity within
the SM. 

The D\O~Collaboration has considered a purely leptophobic model in which 
BR($H^{+}\rightarrow c\bar{s})$=100\%, and the mass of the charged Higgs is close to the
$W$ boson mass. Such a model would decrease the measured $t\bar{t}$ cross-section in the
lepton+jets channel, while leaving the dilepton $t\bar{t}$ cross section unchanged.
Using the ratio between the top pair cross section measured in the dilepton and 
the lepton+jets subsamples, one can then
derive upper bounds on BR($t\rightarrow H^{+}b)$. At 95\% C.L. this upper limit is 
BR($t\rightarrow H^{+}b\rightarrow c\bar{s}b)<$0.35.

The CDF Collaboration has also search for charged Higgs bosons from top decays.
As opposed to the D\O~approach, the CDF analysis seeks to exploit the difference between 
the dijet mass spectra from the regular $W$ decays $W\rightarrow c\bar{s}$ and the charged Higgs decays
$H^{+}\rightarrow c\bar{s}$. Invariant dijet mass templates are built for Higgs masses ranging between
90 and 150 GeV. No significant deviation from the SM is observed, and limits are set for the
BR($t\rightarrow H^{+}b\rightarrow c\bar{s}b)$ as function of the charged Higgs mass. The 95\% C.L. limits 
on this branching ratio vary between 0.32 ($M_{H^{+}}=90$ GeV) and 0.08 ($M_{H^{+}}=130$ GeV).

\section{Conclusions}
We presented selected results from the studies of the properties of the top quarks at the Tevatron.
Both the D\O~and the CDF Collaborations have very mature and diversified top physics programs.
The large amounts of data accumulated by the two experiments allow an unprecedented 
precision for most SM and beyond the SM measurements in the top sector.
For updates of these results and numerous other top-quark-related measurements, we invite the reader to
consult the  D\O~and CDF top group webpages \cite{webpages}.

\begin{acknowledgments}
We thank our D\O~and CDF colleagues for helping us prepare this presentation, and 
to the HCP organizers and participants for providing a very stimulating atmosphere 
throughout the conference.  
\end{acknowledgments}


\begin{thebibliography}{9} 

\bibitem{chgtheo} D.~Chang, W.~Chang, E.~Ma, Phys. Rev. D 59, 091503 (1999).
\bibitem{dcharge} V.~M. Abazov $et$ $al.$ (The D\O~Collaboration), Phys. Rev. Lett. 98, 041801 (2007).
\bibitem{gg} T.~Aaltonen $et$ $al.$ (The CDF Collaboration), arXiv:0807.4262v1 (2008).
\bibitem{d_afb} V.~M.~Abazov $et$ $al.$ (The D\O~Collaboration), 
Phys. Rev. Lett. 100, 142002 (2008).
\bibitem{afb_SM} 
J.~H.~Kuhn and G.~Rodrigo, Phys.\ Rev.\ Lett.\  {\bf 81}, 49 (1998);
J.~H.~Kuhn and G.~Rodrigo, Phys.\ Rev.\  D {\bf 59}, 054017 (1999);
M.~T.~Bowen, S.~Ellis, and D.~Rainwater, Phys. Rev. D 73, 014008 (2006);
S.~Dittmaier, P.~Uwer, and S.~Weinzierl, Phys. Rev. Lett. 98, 262002 (2007);
L.~G.~Almeida, G.~Sterman and W.~Vogelsang, Phys.\ Rev.\  D {\bf 78}, 
014008 (2008).
\bibitem{afb_BSM} 
O.~Antunano, J.~H.~Kuhn and G.~Rodrigo, Phys.\ Rev.\  D {\bf 77}, 014003 (2008);
J.~Rosner, Phys. Lett. B 387, 113 (1996);
P.~Frampton and S.~Glashow, Phys. Lett. B 190, 157 (1987); 
L.~Sehgal and M.~Wanninger, Phys. Lett. B 200, 211 (1988).
\bibitem{jeanny}  T.~Aaltonen $et$ $al.$ (The CDF Collaboration), arXiv:0806.2472v2 (2008).
\bibitem{E6} A.~Leike, Phys. Rept. 317, 143 (1999).
\bibitem{KK} B.~Lillie, L.~Randall, and L.-T.~Wang, JHEP 09, 074 (2007).
T.~G.~Rizzo, Phys. Rev. D 61, 055005 (2000).
\bibitem{topcolor} C.~T.~Hill and S.~Parke, Phys. Rev. D 49, 4454 (1994).
\bibitem{d_nnb} T. Scanlon, FERMILAB-THESIS-2006-43.
\bibitem{marcela} C.~Balazs, M.~Carena, and C.~Wagner, Phys. Rev. D 70, 015007 (2004).
W.~Beenakker $et$ $al.$, Nucl. Phys. B, 515, 3 (1998).
\bibitem{andrew} R.~Erbacher, A.~Ivanov, W.~Johnson, Public CDF note 9343 available at:\\
{\texttt{http://www-cdf.fnal.gov/physics/new/top/2008/tprop/Stop/stopDilCdfNote.pdf}}
\bibitem{d_heli}  V.~M. Abazov $et$ $al.$ (The D\O~Collaboration), Phys. Rev. Lett. 100 , 062004 (2008).
\bibitem{flo}  V.~M. Abazov $et$ $al.$ (The D\O~Collaboration), Phys. Lett. B617, 1 (2005); M.~F.~Canelli, FERMILAB-THESIS-2003-22. 
\bibitem{webpages} D\O~ top quark public webpage:
{\texttt{http://www-d0.fnal.gov/Run2Physics/top/index.html}}; CDF top quark public webpage:
{\texttt{http://www-cdf.fnal.gov/physics/new/top/top.html}}.
\end{thebibliography}
\end{document}